\documentclass[reprint,superscriptaddress,amsmath, amssymb,aps,prd,notitlepage,longbibliography,floatfix,nofootinbib,twocolumn]{revtex4-1}

\usepackage{amsmath}
\allowdisplaybreaks[4]
\usepackage{cancel}
\usepackage{extarrows}
\usepackage{tensor}     
\usepackage{float}
\usepackage[caption = false]{subfig} 
\usepackage[final]{graphicx}   
\usepackage[
colorlinks=true,        
citecolor=blue,         
linkcolor=blue,         
urlcolor=blue           
]{hyperref}             
\usepackage{bm}         
\usepackage{xcolor}     
\usepackage{lipsum}
\usepackage{color}      
\usepackage[utf8]{inputenc} 
\usepackage[section]{placeins} 
\usepackage{appendix}
\usepackage{units}
\usepackage[capitalise]{cleveref}
\usepackage{units}
\newcommand{\nc}{\newcommand*}

\makeatletter
\def\@seccntformat#1{\csname the#1\endcsname.\quad}
\makeatother
\nc{\xbar}{\bar{x}}
\nc{\rhoeq}{\rho_{\mathrm{eq}}}
\nc{\zeq}{z_{\mathrm{eq}}}
\nc{\tla}{\tilde{\lambda}}
\nc{\bt}{\beta}
\nc{\dt}{\delta}
\nc{\Dt}{\Delta}
\nc{\vj}{\vec{j}}
\nc{\vl}{\vec{l}}
\nc{\hx}{\hat{x}}
\nc{\hy}{\hat{y}}
\nc{\bj}{\bm{j}}
\nc{\mJ}{\mathcal{J}}
\nc{\mP}{\mathcal{P}}
\nc{\Msun}{M_\odot}
\nc{\av}[1]{\langle #1 \rangle}
\nc{\eq}[1]{Eq.~\eqref{#1}}
\nc{\al}{\alpha}
\nc{\Xstar}{X_{\ast}}
\nc{\fpbh}{f_{\mathrm{pbh}}}
\nc{\vth}{\vec{\theta}}
\nc{\vla}{\vec{\lambda}}
\nc{\vd}{\vec{d}}
\nc{\Mmin}{M_{\mathrm{min}}}
\nc{\rmd}{\mathrm{d}}
\nc{\mmin}{{m_{\mathrm{min}}}}
\nc{\mmax}{{m_{\mathrm{max}}}}
\nc{\mR}{\mathcal{R}}
\nc{\tmR}{\tilde{\mathcal{R}}}
\nc{\s}{\sigma}
\nc{\ogw}{\Omega_{\mathrm{GW}}}
\nc{\addref}{[\textcolor{red}{add ref}] }
\nc{\Om}{\Omega}
\nc{\gm}{\gamma}
\nc{\Gm}{\Gamma}
\nc{\gpcyr}{\mathrm{Gpc}^{-3}\,\mathrm{yr}^{-1}}
\nc{\Eq}[1]{Eq.~\eqref{#1}}
\nc{\Fig}[1]{Fig.~\ref{#1}}
\nc{\Table}[1]{Table~\ref{#1}}
\nc{\lvc}{LIGO/Virgo} 
\nc{\Sec}[1]{Sec.~\ref{#1}}
\nc{\eg}{\textit{e.g.~}}

\nc{\sovast}{Soviet Ast.}

\begin{document}

\title{Metric Reconstruction and Second Order Perturbation for Generic Spherically Symmetric  spacetime}

\author{Rong-Zhen Guo}
\email{guorongzhen@ucas.ac.cn}
\affiliation{School of Fundamental Physics and Mathematical Sciences, Hangzhou Institute for Advanced Study, UCAS, Hangzhou 310024, China}
\affiliation{School of Physical Sciences, 
    University of Chinese Academy of Sciences, 
    No. 19A Yuquan Road, Beijing 100049, China}

\author{Qing-Guo Huang}
\email{corresponding author: huangqg@itp.ac.cn}
\affiliation{School of Fundamental Physics and Mathematical Sciences, Hangzhou Institute for Advanced Study, UCAS, Hangzhou 310024, China}
\affiliation{School of Physical Sciences, 
    University of Chinese Academy of Sciences, 
    No. 19A Yuquan Road, Beijing 100049, China}
\affiliation{Institute of Theoretical Physics, Chinese Academy of Sciences,Beijing 100190, China}


\begin{abstract}

Higher-order perturbations during the ringdown phase are essential for testing gravitational theories. This requires a perturbation framework that extends beyond General Relativity, as well as an appropriate method for reconstructing the spacetime metric. In this work, we address these challenges within the context of general spherically symmetric spacetimes. We introduce a modified Teukolsky equation for perturbative calculations in asymptotically flat, spherically symmetric spacetimes. The metric reconstruction method, which does not rely on the Hertz potential, is extended to $tr$-symmetric spacetime, allowing for the calculation of metric components under specific gauge conditions. Additionally, we present a second-order perturbation theory applicable to generic spherically symmetric spacetimes.

\end{abstract}
\maketitle

\section{Introduction}

The coalescence of binary black holes (BHs) unfolds in three distinct phases: inspiral, merger, and ringdown. In the inspiral phase, the orbital velocity is much lower than the speed of light, allowing the use of a perturbative post-Newtonian (PN) approximation \cite{Blanchet:2013haa}. During the merger, significant nonlinear effects require numerical relativity for accurate calculations \cite{Bishop:2016lgv}. The ringdown phase, characterized by the emission of gravitational waves (GWs), is well-described by the quasi-normal modes (QNMs) of the final BH \cite{Kokkotas:1999bd, Berti:2009kk}.

QNMs are crucial tools for testing gravitational theories and the nature of BHs. This idea is commonly referred to as BH spectroscopy \cite{Berti:2025hly,Berti:2015itd}. GW250114, which is the loudest GW detected nowadays, has been used to test Einstein's General Relativity (GR) \cite{LIGOScientific:2025obp}. This is mainly because QNMs, especially overtones, are highly sensitive to effects beyond GR. First, the well-established no-hair theorem \cite{Israel:1967za,Israel:1967wq,Carter:1971zc,Hawking:1971vc,Robinson:1975bv} for BHs tells us that the only stationary BH solutions in asymptotically flat 4-dimensional spacetime with known matter fields in GR are the Kerr-Newman family. Therefore, if there are additional unknown fields relative to GR (such as bosonic fields arising from BH superradiant instabilities, or new degrees of freedom in modified gravity theories), or corrections due to quantum effects, they could cause deviations in the actual BH background or the dynamics of perturbations \cite{Blazquez-Salcedo:2016enn,Pani:2011gy,Molina:2010fb,Pani:2011xm}. Second, QNMs can be regarded as `eigenvalues' of GWs on the BH background (albeit in a mathematically informal sense) \cite{Kokkotas:1999bd}. This makes QNMs highly sensitive to boundary conditions. Whether due to BH mimickers or exotic compact objects predicted by certain theories, both could modify the boundary conditions for perturbations \cite{Cardoso:2019rvt, Mark:2017dnq, Barcelo:2017lnx}. Despite the spectral instability, which makes extracting these QNMs from GW signals a complex and challenging task, recent works have attempted to address this issue \cite{Jaramillo:2020tuu,Baibhav:2023clw,Boyanov:2022ark,Crescimbeni:2025ytx}. Recent advancements have significantly improved our understanding of the impact of environmental effects on BH spectroscopy \cite{Cardoso:2025npr,Barausse:2014tra,Cole:2022yzw,Baibhav:2017jhs}.

Most studies on QNMs are based on linear-order BH perturbation theory. However, since gravitational theories are mostly nonlinear, they are more likely to leave imprints in strong-field regions such as BHs. Recent advancements suggest that BH spectroscopy must also account for second-order, quadratic perturbations, which can dominate over linear overtones \cite{Bourg:2024jme,Baibhav:2023clw,London:2014cma,Cheung:2022rbm,Redondo-Yuste:2023seq,Bucciotti:2024zyp}. This implies the existence of what is referred to as quadratic QNMs (QQNMs). Given two linear QNMs $\omega_{\ell_1 \mathrm{~m}_1 \mathrm{n}_1},\omega_{\ell_2 \mathrm{~m}_2 \mathrm{n}_2}$ as seeds, one can find a resulting QQNM $\omega_{QQNM}=\omega_{\ell_1 \mathrm{~m}_1 \mathrm{n}_1}+\omega_{\ell_2 \mathrm{~m}_2 \mathrm{n}_2}$. By carefully examining the relationship between the amplitude of the QQNMs and that of the seed QNMs, one can verify the coupling between QNMs, thereby testing gravitational theories. According to \cite{Yi:2024elj}, Einstein Telescope and Cosmic Explorer could potentially detect QQNMs in several tens of events annually. \cite{Khera:2024bjs} shows that some QQNMs can be observed by space-borne detectors like LISA (Laser Interferometer Space Antenna). This indicates that the study of QQNMs is not only of theoretical interest but also holds the potential for discovery through real GW observations.

These advancements highlight that, just as QNMs are used to test gravitational theories and the nature of BHs, QQNMs can serve the same purpose. Research on QQNMs, in addition to full numerical relativity simulations, can be conducted within the framework of BH perturbation theory \cite{Campanelli:1998jv, Green:2019nam, Ripley:2020xby, Loutrel:2020wbw}. In fact, second-order BH perturbation theory for calculating QQNMs has already been validated within GR \cite{Khera:2024bjs}. To study QQNMs in theories beyond GR, we first need an appropriate linear BH perturbation theory that extends beyond GR. Next, to compute the QQNM couplings, we must calculate the source term, which involves directly obtaining the perturbation of the metric components in the background, which may not be the usual Kerr or Schwarzschild metric. Finally, using second-order perturbation theory, we can compute the corresponding QQNMs. Significant progress has been made in extending linear perturbation theory beyond GR, with some studies focusing on specific cases \cite{Cardoso:2009pk, Manfredi:2017xcv, Moulin:2019ekf, Gogoi:2023kjt, Tattersall:2018nve, Chen:2024mon}, while others have developed more general frameworks \cite{Guo:2024bqe, Li:2022pcy, Guo:2023hdn, Yu:2025wpb}. However, directly computing the metric perturbations in the presence of beyond GR effects remains a challenging task. Even in GR, and for the simple Schwarzschild case, despite a long-standing understanding of its perturbation theory \cite{Regge:1957td, Zerilli:1970se}, calculating the metric perturbations in a given gauge remains an active area of research \cite{Berndtson:2007gsc, Pound:2013faa, Dolan:2012jg}. This challenge constitutes the primary obstacle to using BH perturbation theory to calculate QQNMs.

The topic of this study lies in adapting the well-established procedures in GR, with appropriate modifications, to make them applicable for the calculation of QQNMs in beyond GR scenarios with spherical symmetry. A commonly used procedure in GR is based on the Teukolsky formalism \cite{Teukolsky:1973ha,Press:1973zz,Teukolsky:1974yv}. Although the quantities obtained from the Teukolsky equation are not directly related to the metric components, but instead correspond to gravitational radiation at infinity, the metric information in a given gauge can still be recovered through a method known as metric reconstruction \cite{Whiting:2005hr,Kegeles:1979an,Ori:2002uv,Loutrel:2020wbw,Green:2019nam}. This procedure has been widely applied in various fields, including the study of environmental effects, self-force, and others \cite{Dyson:2025dlj,Dolan:2012jg}. We demonstrate that, by relying on the modified Teukolsky formalism \cite{Guo:2024bqe,Li:2022pcy,Guo:2023hdn}, which remains applicable in beyond GR scenarios, the metric reconstruction procedure provided in \cite{Loutrel:2020wbw} can be directly applied to a broad class of spherically symmetric spacetime, making the calculation of QQNMs possible.

The remainder of this paper is organized as follows. In Sec.\ref{secper}, we briefly review the Teukolsky formalism and demonstrate that the modified Teukolsky equation from \cite{Guo:2024bqe} has broader applicability, being valid for the most general spherically symmetric spacetime. In Sec.\ref{sec}, we show that under suitable gauge choices, the metric procedure can be directly used in general spherically symmetric spacetime with $tr$-symmetry. In Sec.\ref{secsecond}, we show that by fixing the gauge order by order, the linear-order modified Teukolsky equation can be directly extended to second-order perturbation theory, with source terms that include contributions identical to those in GR. The conclusion and outlook are presented in Sec.\ref{conc}. Unless otherwise stated, we adopt geometric units with \( c = G = 1 \) throughout this paper. Note that, since \(\pi\) refers to a specific scalar in the NP formalism, the variant \(\varpi\) is used in this paper to represent the normal meaning of \(\pi\).

\medskip

\section{Perturbative Formalism} \label{secper}
\subsection{The Brief Review of NP Formalism}
In NP formalism, we choose four null vectors \( l^\mu, n^\mu, m^\mu, \bar{m}^\mu \) as a tetrad, where \( l^\mu, n^\mu \) are real-valued vectors and \( m^\mu, \bar{m}^\mu \) are complex-valued vectors. These vectors satisfy the following conditions:
\begin{equation}
\begin{aligned}
l_\mu l^\mu = n_\mu n^\mu & = m_\mu m^\mu = \bar{m}_\mu \bar{m}^\mu = 0, \\
l_\mu n^\mu & = -m_\mu \bar{m}^\mu = 1, \\
l_\mu m^\mu = l_\mu \bar{m}^\mu & = n_\mu m^\mu = n_\mu \bar{m}^\mu = 0,
\end{aligned}
\label{orth}
\end{equation}
where the bar represents the complex conjugate. In this tetrad, the metric can be written as:
\begin{equation}
g_{\mu \nu} = l_\mu n_\nu + n_\mu l_\nu - m_\mu \bar{m}_\nu - \bar{m}_\mu m_\nu.
\end{equation}

As is commonly adopted in the literature, we introduce the directional derivatives:
\begin{equation}
\begin{aligned}
    D &\equiv l^\mu \nabla_\mu, \\
    \Delta &\equiv n^\mu \nabla_\mu, \\
    \delta &\equiv m^\mu \nabla_\mu, \\
    \bar{\delta} &\equiv \bar{m}^\mu \nabla_\mu,
\end{aligned}
\end{equation}
where \( \nabla_\mu \) is the covariant derivative compatible with the spacetime metric \( g_{\mu \nu} \).

The key variables in NP formalism consist of 5 Weyl scalars \( \left( \Psi_1, \Psi_2, \dots \right) \), 12 spin coefficients \( (\kappa, \pi, \varepsilon, \dots) \), and 10 NP Ricci scalars \( (\Phi_{00}, \Phi_{01}, \dots, \Lambda) \). The definitions for these quantities are provided in Appendix.~\ref{NP}. These variables enable the derivation of fundamental NP relations, including 18 complex Ricci identities and 9 complex as well as 2 real Bianchi identities. In this study, we use the following three equations:

\begin{widetext}
\begin{equation}
\begin{aligned}
D \Psi_1 - \bar{\delta} \Psi_0 - D \Phi_{01} + \delta \Phi_{00} &= (\pi - 4 \alpha) \Psi_0 + 2(2 \rho + \epsilon) \Psi_1 - 3 \kappa \Psi_2 - (\bar{\pi} - 2 \bar{\alpha} - 2 \beta) \Phi_{00} \\
& \quad - 2(\bar{\rho} + \epsilon) \Phi_{01} - 2 \sigma \Phi_{10} + 2 \kappa \Phi_{11} + \bar{\kappa} \Phi_{02}, \\
\Delta \Psi_0 - \delta \Psi_1 + D \Phi_{02} - \delta \Phi_{01} &= (4 \gamma - \mu) \Psi_0 - 2(2 \tau + \beta) \Psi_1 + 3 \sigma \Psi_2 - \bar{\lambda} \Phi_{00} \\
& \quad + 2(\bar{\pi} - \beta) \Phi_{01} + 2 \sigma \Phi_{11} + (\bar{\rho} + 2 \epsilon - 2 \bar{\epsilon}) \Phi_{02} - 2 \kappa \Phi_{12}, \\
D \sigma - \delta \kappa &= (\rho + \bar{\rho} + 3 \epsilon - \bar{\epsilon}) \sigma - (\tau - \bar{\pi} + \bar{\alpha} + 3 \beta) \kappa + \Psi_0.
\end{aligned}
\label{equ1}
\end{equation}
\end{widetext}

\subsection{The Modified Teukolsky Equation and Its Regime of Validity}
Consider a four-dimensional, asymptotically flat, spherically symmetric static metric, which can generally be written as
\[
\mathrm{d}s^2 = G^2(r) \, \mathrm{d}t^2 - \frac{1}{F^2(r)} \, \mathrm{d}r^2 - r^2 \left( \mathrm{d}\theta^2 + \sin^2 \theta \, \mathrm{d}\varphi^2 \right), \label{metric1}
\]
where the asymptotic flatness conditions are given by
\[
F(r) \underset{r \to +\infty}{\longrightarrow} 1, \quad G(r) \underset{r \to +\infty}{\longrightarrow} 1.
\]

If we further assume that
\[
G^2(r) = F^2(r),
\]
then the spacetime is called \( t r \)-symmetric. It is straightforward to verify that this spacetime must be Type-D, which implies that:
\begin{equation}
\begin{gathered}
\Psi_0^{(0)} = \Psi_1^{(0)} = \Psi_3^{(0)} = \Psi_4^{(0)} = 0, \\
\kappa^{(0)} = \sigma^{(0)} = \nu^{(0)} = \lambda^{(0)} = 0.
\end{gathered}
\end{equation}
The scalar with the upper index \({}^{(n)}\) represents the \(n\)-th order perturbation of this scalar (For simplicity, background quantities will no longer be denoted with superscripts).

We choose the null tetrad as:
\begin{equation}
\begin{aligned}
l^\mu &= \left( \frac{1}{G^2(r)}, \frac{F(r)}{G(r)}, 0, 0 \right), & n^\mu &= \frac{1}{2} \left( 1, -G(r) F(r), 0, 0 \right), \\
m^\mu &= \left( 0, 0, \frac{1}{\sqrt{2} r}, \frac{i \csc \theta}{\sqrt{2} r} \right), & \bar{m}^\mu &= \left( 0, 0, \frac{1}{\sqrt{2} r}, -\frac{i \csc \theta}{\sqrt{2} r} \right).
\end{aligned}
\end{equation}
The non-zero NP scalars in this tetrad are:
\[
\rho, \, \mu, \, \gamma, \, \alpha, \, \beta, \, \Psi_2, \, \Lambda, \, \Phi_{11}, \, \Phi_{00}, \, \Phi_{22}.
\]
Thus, Eq.~(\ref{equ1}) can be expressed as:
\begin{widetext}
\begin{equation}
\begin{aligned}
& (\bar{\delta} - 4 \alpha + \pi) \Psi_0^{(1)} - (D - 4 \rho - 2 \epsilon) \Psi_1^{(1)} - 3 \kappa^{(1)} \Psi_2 + 2 \kappa^{(1)} \Phi_{11} - (\delta + \bar{\pi} - 2 \bar{\alpha} - 2 \beta)^{(1)} \Phi_{00} = S_1, \\
& (\Delta - 4 \gamma + \mu) \Psi_0^{(1)} - (\delta - 4 \tau - 2 \beta) \Psi_1^{(1)} - 3 \sigma^{(1)} \Psi_2 - 2 \sigma^{(1)} \Phi_{11} + \bar{\lambda}^{(1)} \Phi_{00} = S_2, \\
& (D - \rho - \bar{\rho} - 3 \epsilon + \bar{\epsilon}) \sigma^{(1)} - (\delta - \tau + \bar{\pi} - \bar{\alpha} - 3 \beta) \kappa^{(1)} - \Psi_0^{(1)} = 0.
\end{aligned}
\label{use}
\end{equation}
\end{widetext}

Following \cite{Guo:2024bqe} (we support a quick review of this procedure in Appendix.~\ref{Mypaper} ), it can be observed that when the metric satisfies the so-called \( t r \)-symmetric condition, the equation can ultimately be reduced to a decoupled Teukolsky-like equation. This equation uses \( \sigma^{(1)} = 0 \) as a gauge-fixing condition. Moreover, the derivation makes full use of the spherical symmetry of the background metric. Compared to the vacuum solution in GR, the main modifications arise from the source term introduced by deviations from GR and the presence of NP scalars due to the background's non-Ricci-flatness. 

In a more general case, additional terms can be found compared to the previously discussed scenario, and they are given by:
\begin{equation}
(\delta + \bar{\pi} - 2 \bar{\alpha} - 2 \beta)^{(1)} \Phi_{00}; \quad \bar{\lambda}^{(1)} \Phi_{00}.
\end{equation}

We demonstrate that in this scenario, it is always possible to impose the following gauge condition, which can be achieved through the gauge freedom introduced by the tetrad transformation. These conditions do not lead to any contradictions:
\begin{equation}
    \sigma^{(1)} = 0, \quad \lambda^{(1)} = 0, \quad (\delta + \bar{\pi} - 2 \bar{\alpha} - 2 \beta)^{(1)} = 0.
\end{equation}

In the NP formalism, there are three types of transformations that preserve the orthogonality conditions in Eq.~(\ref{orth}):

\begin{equation}
\begin{aligned}
\mathrm{I}: & \quad l \rightarrow l, \\
           & \quad m \rightarrow m + a l, \\
           & \quad \bar{m} \rightarrow \bar{m} + \bar{a} l, \\
           & \quad n \rightarrow n + \bar{a} m + a \bar{m} + a \bar{a} l, \\
\mathrm{II}: & \quad n \rightarrow n, \\
            & \quad m \rightarrow m + b n, \\
            & \quad \bar{m} \rightarrow \bar{m} + \bar{b} n, \\
            & \quad l \rightarrow l + \bar{b} m + b \bar{m} + b \bar{b} n, \\
\mathrm{III}: & \quad l \rightarrow A^{-1} l, \\
             & \quad n \rightarrow A n, \\
             & \quad m \rightarrow e^{i \theta} m, \\
             & \quad \bar{m} \rightarrow e^{-i \theta} \bar{m}.
\end{aligned}
\end{equation}

Note that Eq.~(\ref{use}) is derived from the gauge-invariant Bianchi and Ricci identities. Therefore, under gauge transformations, the form of Eq.~(\ref{use}) itself remains unchanged.

Here, we list the relations of the NP scalars under the three relevant types of transformation. For type-I case,
\begin{equation}
\begin{aligned}
\sigma & \mapsto \sigma+a \kappa, \\
\beta & \mapsto \beta+a \epsilon+\bar{a} \sigma+a \bar{a} \kappa, \\
\lambda & \mapsto \lambda+\bar{a} \pi+2 \bar{a} \alpha+2 \bar{a}^2 \epsilon+\bar{a}^2 \rho\\
&\quad+\bar{a}^3 \kappa+\bar{\delta} \bar{a}+\bar{a} D \bar{a}, \\
\alpha & \mapsto \alpha+\bar{a} \epsilon+\bar{a} \rho+\bar{a}^2 a \kappa, \\
\pi & \mapsto \pi+2 \bar{a} \epsilon+\bar{a}^2 a \kappa+D \bar{a} .
\end{aligned}
\end{equation}
For type-II case,
\begin{equation}
\begin{aligned}
\pi & \mapsto \pi+b \lambda+\bar{b} \mu+b \bar{b} \nu, \\
\lambda & \mapsto \lambda+\bar{b} \nu, \\
\alpha & \mapsto \alpha+\bar{b} \tau+b \lambda+b \bar{b} \nu, \\
\beta & \mapsto \beta+b \gamma+b \mu+b^2 \nu, \\
\sigma &\mapsto \sigma+b \tau+2 b \beta+2 b^2 \gamma\\
&+b^2 \mu+b^3 \nu-\delta b-b \Delta b, .
\end{aligned}
\end{equation}
For type-III case,
\begin{equation}
\begin{aligned}
\pi & \mapsto e^{-i \theta} \pi, \\
\alpha & \mapsto \frac{1}{2} e^{-i \theta} \left[ \frac{1}{A} \bar{\delta} A + e^{-i \theta} \bar{\delta} e^{i \theta} + 2 \alpha \right], \\
\sigma & \mapsto A e^{2 i \theta} \sigma, \\
\beta & \mapsto \frac{1}{2} e^{-i \theta} \left[ \frac{1}{A} \delta A + e^{-i \theta} \delta e^{i \theta} + 2 \beta \right], \\
\lambda & \mapsto \frac{1}{A} e^{-2 i \theta} \lambda.
\end{aligned}
\end{equation}

By utilizing the fact that the spacetime itself is of Type-D, it is straightforward to observe that \( \sigma^{(1)} \) is invariant under Type-I gauge transformation (i.e., infinitesimal Type-I tetrad transformation), and \( \lambda^{(1)} \) is invariant under Type-II gauge transformation. This implies that there always exists an appropriate Type-I transformation such that \( \lambda^{(1)} = 0 \), and a Type-II transformation that ensures \( \sigma^{(1)} = 0 \), with both conditions being satisfiable simultaneously.

After performing such a gauge transformation, both \( \lambda^{(1)} \) and \( \sigma^{(1)} \) become gauge-invariant quantities under Type-III transformations. However, since \( \alpha^{(1)}, \beta^{(1)}, \pi^{(1)} \) are not invariant under Type-III transformations, one can use the residual gauge freedom to set
\begin{equation}
    (\delta + \bar{\pi} - 2 \bar{\alpha} - 2 \beta)^{(1)} = 0.
\end{equation}

In conclusion, by fully utilizing the gauge degrees of freedom, we observe that the most general form of Eq. (\ref{use}) reduces to the same form as in \cite{Guo:2024bqe}. Using the same derivation, it can be shown that the decoupled modified Teukolsky equation in \cite{Guo:2024bqe} applies to the most general form of four-dimensional asymptotically flat spherically symmetric spacetime and is given by:
\begin{equation}
\begin{aligned}
& \left[ (D - 3 \epsilon + \bar{\epsilon} - 4 \rho - \bar{\rho})(\Delta - 4 \gamma + \mu) \right. \\
& \left. - (\delta + \bar{\pi} - \bar{\alpha} - 3 \beta - 4 \tau)(\bar{\delta} + \pi - 4 \alpha) \right. \\
& \left. - 3 \Psi_2 + 2 \Phi_{11} \right] \Psi_0^{(1)} = 4 \varpi T_0, \\
& \left[ (\Delta + 3 \gamma - \bar{\gamma} + 4 \mu + \bar{\mu})(D + 4 \epsilon - \rho) \right. \\
& \left. - (\bar{\delta} - \bar{\tau} + \bar{\beta} + 3 \alpha + 4 \pi)(\delta - \tau + 4 \beta) \right. \\
& \left. - 3 \Psi_2 + 2 \Phi_{11} \right] \Psi_4^{(1)} = 4 \varpi T_4.
\end{aligned}
\label{TeukEq}
\end{equation}
where \( T_0 \) and \( T_4 \) are the source terms that have a similar form in GR. The second equation for \( \Psi_4^{(1)} \), which is directly related to the GWs at null infinity, is derived by taking a Geroch-Held-Penrose transformation:
\begin{equation}
l^\mu \mapsto n^\mu, \quad n^\mu \mapsto l^\mu, \quad m^\mu \mapsto \bar{m}^\mu, \quad \bar{m}^\mu \mapsto m^\mu,
\end{equation}
which keeps the NP formalism invariant.

It is important to emphasize that, since \( \Psi_0^{(1)} \) and \( \Psi_4^{(1)} \) are gauge-invariant quantities for Type-D spacetime, the results obtained in our specific gauge are themselves gauge-invariant. As shown in \cite{Nakajima:2024qrq, Guo:2024bqe}, the apparent inconsistencies in the modified Teukolsky formalism \cite{Nakajima:2024qrq, Guo:2024bqe, Li:2022pcy} commonly arise from the impact of gauge choices on the source term, and therefore do not imply any contradiction.

\section{Metric Reconstruction for $tr$-symmetric metrics}\label{sec}
In this section, we demonstrate that the metric reconstruction procedure for GR vacuum solutions, as presented in \cite{Loutrel:2020wbw}, can be directly extended to the spherical \( t r \)-symmetric metrics. This extension allows for the computation of source terms that are essential for evaluating second-order perturbations.

\subsection{Review of the Procedure in Kerr Case}
In this section, we demonstrate that the metric reconstruction procedure for GR vacuum solutions, as presented in \cite{Loutrel:2020wbw}, can be directly extended to spherical \( t r \)-symmetric metrics. This extension allows for the computation of source terms necessary for evaluating second-order perturbations.

The metric reconstruction procedure is crucial in studies based on BH perturbation theory. Whether for higher-order QNMs, the dynamics of GWs and other fields, or the self-force problem, a precise understanding of gravitational perturbations in a curved spacetime background under a given gauge is essential. Currently, there are two methods for metric reconstruction using the NP invariant scalars \(\Psi_0^{(1)}\) or \(\Psi_4^{(1)}\), differing in the use of the well-known Hertz potential \cite{Wald:1973wwa}. Both methods have their respective advantages, depending on the application. In this work, we focus on the approach outlined in \cite{Loutrel:2020wbw}, which directly constructs the metric perturbation using \(\Psi_4^{(1)}\) without requiring the Hertz potential.

The procedure is as follows (for further details, see Appendix \ref{procedure}):
\begin{itemize}
    \item Compute \(\Psi_0^{(1)}\) or \(\Psi_4^{(1)}\) using the Teukolsky equation (we will use \(\Psi_4^{(1)}\) as an example);
    \item Choose a radiation gauge. The gauge choice involves two parts. The first part defines the relationship between the first-order metric perturbation and the first-order tetrad perturbation. The first-order perturbation of the tetrad can be expanded in terms of the local, background-fixed zeroth-order tetrad:

    \[
    \left( \begin{array}{c}
    l_\mu^{(1)} \\
    n_\mu^{(1)} \\
    m_\mu^{(1)} \\
    \bar{m}_\mu^{(1)}
    \end{array} \right)
    =
    \left( \begin{array}{cccc}
    b_{11} & b_{12} & c_{13} & \bar{c}_{13} \\
    b_{21} & b_{22} & c_{23} & \bar{c}_{23} \\
    c_{31} & c_{32} & c_{33} & c_{34} \\
    \bar{c}_{31} & \bar{c}_{32} & \bar{c}_{34} & \bar{c}_{33}
    \end{array} \right)
    \left( \begin{array}{c}
    l_\mu^{(0)} \\
    n_\mu^{(0)} \\
    m_\mu^{(0)} \\
    \bar{m}_\mu^{(0)}
    \end{array} \right)
    \]
    where the \(b_{ij}\) are real coefficients and the \(c_{ij}\) are complex coefficients. Following \cite{Loutrel:2020wbw, Campanelli:1998jv, Chrzanowski:1976jy}, we use the six degrees of freedom of the linearized tetrad vectors to select \(b_{11} = c_{13} = c_{23} = \operatorname{Im}(c_{33}) = 0\). The second part involves the choice of radiation gauge, either the ingoing radiation gauge (IRG) or the outgoing radiation gauge (ORG) (we use ORG as an example; both types will be discussed in detail below);
    \item \(\Psi_4^{(1)}\) directly determines \(\lambda^{(1)}\), which in turn determines the metric component \(h_{mm}\) in the null tetrad;
    \item Using \(\Psi_4^{(1)}\), compute \(\psi_3^{(1)}\), which directly determines \(\pi^{(1)}\), and consequently determines \(h_{lm}\);
    \item Using \(\Psi_4^{(1)}\), compute \(\psi_2^{(1)}\) to obtain \(h_{mm}\). At this point, all of the non-zero metric components in ORG have been determined, completing the metric reconstruction.
\end{itemize}

This procedure can be described using the Geroch-Held-Penrose transformation in the NP formalism:

\[
l^\mu \mapsto n^\mu, \quad n^\mu \mapsto l^\mu, \quad m^\mu \mapsto \bar{m}^\mu, \quad \bar{m}^\mu \mapsto m^\mu,
\]
which yields the dual counterpart (i.e., \(\psi_4^{(1)} \leftrightarrow \psi_0^{(1)}\), ORG \(\leftrightarrow\) IRG). This method is similar to the approach discussed earlier in this study, where the metric is constructed by fully exploiting gauge conditions to simplify the inherently complex NP equations, leading to a set of differential equations that facilitate the metric computation. The ORG condition can be expressed as:

\[
n^\mu h_{\mu\nu} = 0.
\]

Due to the presence of additional residues in the gauge choice for type-D spacetime, we can further impose the traceless condition:

\[
h_{\mu}^{\mu} = g^{\mu\nu} h_{\mu\nu} = 0.
\]

Finally, we translate all of the aforementioned gauge conditions into the language of NP formalism, expressing them using NP scalars and metric components under the null tetrad:

\[
h_{ln} = h_{nn} = h_{nm} = h_{n\bar{m}} = h_{m\bar{m}} = 0;
\]

\[
\nu^{(1)} = \mu^{(1)} = \gamma^{(1)} = \Phi_{22}^{(1)} = 0.
\label{gaugeuse}
\]

The construction procedure involves the following NP equations \cite{Loutrel:2020wbw} (note that the equations presented here are in their original, unperturbed form):

\begin{equation}
\Delta \lambda - \bar{\delta} \nu = -\lambda(\mu + \bar{\mu} + 3 \gamma - \bar{\gamma}) + \nu(3 \alpha + \bar{\beta} + \pi - \bar{\tau}) - \Psi_4,
\label{A9j}
\end{equation}

\begin{equation}
\lambda = \frac{1}{2} \left[ -\Delta + 2(\bar{\gamma} - \gamma) + \mu - \bar{\mu} \right] h_{\bar{m} \bar{m}} - (\pi + \bar{\tau}) h_{n \bar{m}},
\label{c1a}
\end{equation}

\begin{equation}
- \Delta \Psi_3 + \delta \Psi_4 + 3 \nu \Psi_2 - 2(\gamma + 2 \mu) \Psi_3 - (\tau - 4 \beta) \Psi_4 + \mathcal{R}_h = 0,
\label{A10h}
\end{equation}

\begin{equation}
\begin{aligned}
\mathcal{R}_h = & \Delta \Phi_{21} - \bar{\delta} \Phi_{22} + 2(\bar{\mu} + \gamma) \Phi_{21} - 2 \nu \Phi_{11} - \bar{\nu} \Phi_{20} \\
& + 2 \lambda \Phi_{12} + (\bar{\tau} - 2 \alpha - 2 \bar{\beta}) \Phi_{22},
\end{aligned}
\label{Rh}
\end{equation}

\begin{equation}
D \nu - \Delta \pi = \mu(\pi + \bar{\tau}) + \lambda(\bar{\pi} + \tau) + \pi(\gamma - \bar{\gamma}) - \nu(3 \epsilon + \bar{\epsilon}) + \Psi_3 + \Phi_{21},
\label{A9i}
\end{equation}

\begin{equation}
\begin{aligned}
\pi = & -\frac{1}{2} (D + 2 \epsilon - \rho) h_{n \bar{m}} - \frac{1}{2} (\Delta - 2 \bar{\gamma} + \bar{\mu}) h_{l \bar{m}} \\
& + \frac{1}{2} (\bar{\delta} - \pi - \bar{\tau}) h_{l n} - \frac{1}{2} \bar{\tau} h_{m \bar{m}} - \frac{1}{2} \tau h_{\bar{m} \bar{m}},
\end{aligned}
\label{C1l}
\end{equation}

\begin{equation}
\begin{aligned}
\tau = & \frac{1}{2} (D + 2 \bar{\epsilon} - \bar{\rho}) h_{n m} + \frac{1}{2} (\Delta - 2 \gamma + \mu) h_{l m} \\
& - \frac{1}{2} (\delta + \bar{\pi} + \tau) h_{l n} - \frac{1}{2} \pi h_{m m} - \frac{1}{2} \bar{\pi} h_{m \bar{m}},
\end{aligned}
\label{C1k}
\end{equation}

\begin{equation}
- \Delta \Psi_2 + \delta \Psi_3 + 2 \nu \Psi_1 - 3 \mu \Psi_2 + 2(\beta - \tau) \Psi_3 + \sigma \Psi_4 + \mathcal{R}_g = 0,
\label{A10g}
\end{equation}

\begin{equation}
\begin{aligned}
\mathcal{R}_g = & - D \Phi_{22} + \delta \Phi_{21} + 2(\bar{\pi} + \beta) \Phi_{21} - 2 \mu \Phi_{11} - \bar{\lambda} \Phi_{20} \\
& + 2 \pi \Phi_{12} + (\bar{\rho} - 2 \epsilon - 2 \bar{\epsilon}) \Phi_{22} - 2 \Delta \Lambda,
\end{aligned}
\label{Rg}
\end{equation}

\begin{equation}
\begin{aligned}
D \gamma - \Delta \epsilon = & \alpha(\tau + \bar{\pi}) + \beta(\bar{\tau} + \pi) \\
& - \gamma(\epsilon + \bar{\epsilon}) - \epsilon(\gamma + \bar{\gamma}) + \tau \pi \\
& - \nu \kappa + \Psi_2 + \Phi_{11} - \Lambda,
\end{aligned}
\label{A9f}
\end{equation}

\begin{equation}
\begin{aligned}
\epsilon = & \frac{1}{4} \left( -\Delta + 2 \bar{\gamma} + \mu - \bar{\mu} \right) h_{l l} + \frac{1}{4} \left( 2 D + \rho - \bar{\rho} \right) h_{l n} \\
& + \frac{1}{4} \left( -\delta + 2 \bar{\alpha} - \bar{\pi} - 2 \tau \right) h_{l \bar{m}} \\
& + \frac{1}{4} \left( \bar{\delta} - 2 \alpha - 3 \pi - 2 \bar{\tau} \right) h_{l m} \\
& + \frac{1}{4} \left( \rho - \bar{\rho} \right) h_{m \bar{m}}.
\end{aligned}
\label{C1g}
\end{equation}

The specific method and details for reconstructing the metric perturbations in the Kerr background using these equations can be found in \cite{Loutrel:2020wbw}, with a brief review provided in the appendix.

After applying the first-order perturbation to the equations listed above in the non-vacuum \( t r \)-symmetric case, one can see that non-Ricci-flatness only affects Eqs. (\ref{A10h}) and (\ref{A10g}). These two equations are essentially Bianchi identities, which naturally lead to the involvement of the Ricci tensor. According to Eqs. (\ref{Rh}) and (\ref{Rg}), the contributions from non-zero \(\Phi_{ij}^{(0)}\) are:

\begin{equation}
\mathcal{R}_g: -2 \mu^{(1)} \Phi_{11}; \quad \mathcal{R}_h: -2 \nu^{(1)} \Phi_{11}.
\end{equation}

However, in the traceless ORG case, the gauge condition in Eq. (\ref{gaugeuse}) cancels all these terms. This implies that in a \( t r \)-symmetric spherically symmetric spacetime, the metric perturbation can be reconstructed in a manner analogous to GR. Specifically, one can compute or estimate the contributions analogous to the stress-energy tensor in GR. Subsequently, the same method can be applied to solve a set of first-order differential equations, ultimately obtaining the metric components within the appropriate regions.

The only point to note is the constraint \(\Phi_{22}^{(1)} = 0\). In fact, even in GR, this condition is challenging to satisfy in calculations such as the self-force in EMRI. This is one of the drawbacks of methods that do not rely on the Hertz potential. Nevertheless, considering that the inhomogeneous metric reconstruction in GR remains complex, this approach is still valuable, particularly in time-domain calculations of the second-order ringdown waveform.

\medskip
\section{Second Order Perturbations}\label{secsecond}
When dealing with higher-order perturbation equations, the derivation of the modified Teukolsky formalism, i.e., Eq. (\ref{equ1}), leads to the following equation \cite{Campanelli:1998jv}:

\begin{widetext}
\begin{equation}
\begin{aligned}
& (\bar{\delta} - 4 \alpha + \pi) \Psi_0^{(n)} - (D - 4 \rho - 2 \epsilon) \Psi_1^{(n)} - 3 \kappa^{(n)} \Psi_2 + 2 \kappa^{(n)} \Phi_{11} - (\delta + \bar{\pi} - 2 \bar{\alpha} - 2 \beta)^{(n)} \Phi_{00} = S_1^{(n)} + T_1^{(n)}, \\
& (\Delta - 4 \gamma + \mu) \Psi_0^{(n)} - (\delta - 4 \tau - 2 \beta) \Psi_1^{(n)} - 3 \sigma^{(n)} \Psi_2 - 2 \sigma^{(n)} \Phi_{11} + \bar{\lambda}^{(n)} \Phi_{00} = S_2^{(n)} + T_2^{(n)}, \\
& (D - \rho - \bar{\rho} - 3 \epsilon + \bar{\epsilon}) \sigma^{(n)} - (\delta - \tau + \bar{\pi} - \bar{\alpha} - 3 \beta) \kappa^{(n)} - \Psi_0^{(n)} = T_3^{(n)}.
\end{aligned}
\label{usehigher}
\end{equation}
\end{widetext}

Here, \( S_i^{(n)} \) represents the contribution from \( \Phi_{ij} \)'s, while \( T_i^{(n)} \) corresponds to the contribution from the metric components, Weyl scalars, and other spin coefficients. In general, higher-order Weyl scalars are no longer gauge-invariant. However, due to the peeling theorem \cite{Wald:1984rg} in asymptotically flat spacetime, \( \Psi_4 \) provides a non-perturbative relation to gravitational radiation at null infinity. This relationship explains why this method is frequently employed to calculate higher-order perturbations in black hole perturbation theory, such as second-order QNMs.

Compared to leading-order perturbations, the main difficulty with higher-order calculations arises from the fact that many spin coefficients, which were originally invariant under the three types of gauge transformations, become gauge-dependent. This suggests that the previously mentioned tricks may no longer be directly applicable. However, by employing the metric reconstruction method discussed in this paper, at least in the case of second-order perturbations, one can still derive a second-order equation that is formally identical to the first-order perturbation equation. The only difference is that the source term now contains contributions identical in form to those found in the second-order perturbation theory of classical GR.

The key point here lies in the choice of gauge during the metric reconstruction. As discussed in Sec.\ref{sec}, the six first-order gauge degrees of freedom induced by the frame have been completely determined. This implies that when discussing the gauge transformations of NP scalars at second-order perturbations, with the first-order metric fixed, it is no longer possible to introduce corrections arising from first-order gauge transformations and first-order perturbed NP scalars.

Let's now proceed with a more detailed calculation. At second-order, the form of \( T_i^{(2)} \)'s can be obtained as follows:

\begin{widetext}
\begin{equation}
\begin{aligned}
& (\bar{\delta} - 4 \alpha + \pi)^{(1)} \Psi_0^{(1)} - (D - 4 \rho - 2 \epsilon)^{(1)} \Psi_1^{(1)} - 3 \kappa^{(1)} \Psi_2^{(1)} = -T_1^{(2)}, \\
& (\Delta - 4 \gamma + \mu)^{(1)} \Psi_0^{(1)} - (\delta - 4 \tau - 2 \beta)^{(1)} \Psi_1^{(1)} - 3 \sigma^{(1)} \Psi_2^{(1)} = -T_2^{(2)}, \\
& (D - \rho - \bar{\rho} - 3 \epsilon + \bar{\epsilon})^{(1)} \sigma^{(1)} - (\delta - \tau + \bar{\pi} - \bar{\alpha} - 3 \beta)^{(1)} \kappa^{(1)} = -T_3^{(2)}.
\end{aligned}
\end{equation}
\end{widetext}

It is clear that after performing a calculation similar to the one in \cite{Guo:2024bqe}, the resulting source term matches exactly with the one derived in \cite{Campanelli:1998jv}. This is because both methods are mathematically inspired by the derivation of the original Teukolsky equation \cite{Teukolsky:1973ha}. Both use operator identities of the form:

\begin{equation}
\begin{aligned}
& \left[ D - (p+1) \epsilon + \bar{\epsilon} + q \rho - \bar{\rho} \right] (\delta - p \beta + q \tau) \\
& - \left[ \delta - (p+1) \beta - \bar{\alpha} + \bar{\pi} + q \tau \right] (D - p \epsilon + q \rho) = 0.
\end{aligned}
\label{identity}
\end{equation}

Following the steps in \cite{Guo:2024bqe} (and see Appendix.\ref{Mypaper}), one naturally arrives at a sourced differential equation, with the left-hand side being completely identical to Eq. (\ref{TeukEq}). Since the only allowed gauge transformations are those purely due to the second-order transformation of the background tetrad, the gauge-fixing procedure discussed earlier can once again be applied, simplifying the NP equations.

\medskip

\section{Discussions and Conclusions}\label{conc}

This work introduces a modified formalism for calculations in BH perturbation theory, which is applicable to a broad class of spacetime. Our findings can be summarized in three key points.

First, the modified Teukolsky equation developed in this study extends the classical Teukolsky framework, offering broader applicability to perturbative analyses in any asymptotically flat, spherically symmetric spacetime. Notably, this formalism is also effective in special cases of deformed Kerr spacetime, as demonstrated in modified gravity theories such as those explored in \cite{Guo:2023wtx}. The versatility of this formalism makes it a valuable tool for various perturbative studies, including QNMs, self-force calculations, and the dynamics of other fields in curved spacetime.

Second, for \( t r \)-symmetric spacetimes, we show that the metric reconstruction procedure, which avoids the use of the Hertz potential, can be directly extended to these cases. This extension enables the calculation of metric components in the radiation gauge, which is widely used in GR. This simplification is particularly useful in the study of topics such as QQNMs in modified gravity theories, offering a more tractable framework for these studies.

Third, by systematically fixing the gauge degrees of freedom induced by the tetrad transformation at each order, we demonstrate that the modified Teukolsky formalism naturally generalizes to second-order perturbation theory. This extension mirrors the treatment in classical GR, while incorporating additional source terms from the \( \Phi_{ij} \) components introduced by alternative theories of gravity. Moreover, the source terms in our formalism are consistent with those in second-order perturbation theory for type-D spacetimes in GR, providing a solid foundation for further studies in modified gravity.

Despite these advancements, there are limitations to the formalism presented here. In particular, for certain modified gravity theories, the source terms cannot be fully derived from the NP formalism alone. While the NP formalism is powerful for describing the geometry of spacetime, it does not capture the complete dynamics of gravitational fields in alternative theories, which must be computed within the broader framework of field theory. Therefore, future research will be needed to extend this formalism to incorporate the dynamical equations of modified gravity theories, with additional studies comparing our approach to existing work on alternative gravity models \cite{Li:2025fci, Li:2022pcy}.

That being said, there are models particularly well-suited for study within the framework developed in this work. One such example is the Quantum Oppenheimer-Snyder (QOS) model introduced in \cite{Lewandowski:2022zce}, which combines a loop quantum gravity (LQG) description of the black hole interior with an external GR solution via matching conditions. In this context, perturbations outside the BH can be analyzed under the source-free homogeneous assumption, making it an ideal candidate for further exploration within the formalism we have presented. We hope that this study contributes to the expanding body of work on BH perturbation theory with quantum corrections and look forward to future developments in this exciting area of research.

\textit{Acknowledgements.}
QGH is supported by the grants from NSFC (Grant No.~12547110, 12475065, 12447101) and the China Manned Space Program with grant no. CMS-CSST-2025-A01.

\begin{appendices}
\section{Defination of NP Scalars}\label{NP}
In the NP formalism, the 12 complex spin coefficients can be defined as:
\begin{equation}
\begin{aligned}
& \kappa := -m^a D \ell_a = -m^a \ell^b \nabla_b \ell_a, \quad \tau := -m^a \Delta \ell_a = -m^a n^b \nabla_b \ell_a, \\
& \sigma := -m^a \delta \ell_a = -m^a m^b \nabla_b \ell_a, \quad \rho := -m^a \bar{\delta} \ell_a = -m^a \bar{m}^b \nabla_b \ell_a; \\
& \pi := \bar{m}^a D n_a = \bar{m}^a \ell^b \nabla_b n_a, \quad \nu := \bar{m}^a \Delta n_a = \bar{m}^a n^b \nabla_b n_a, \\
& \mu := \bar{m}^a \delta n_a = \bar{m}^a m^b \nabla_b n_a, \quad \lambda := \bar{m}^a \bar{\delta} n_a = \bar{m}^a \bar{m}^b \nabla_b n_a; \\
& \varepsilon := -\frac{1}{2} \left( n^a D \ell_a - \bar{m}^a D m_a \right) = -\frac{1}{2} \left( n^a \ell^b \nabla_b \ell_a - \bar{m}^a \ell^b \nabla_b m_a \right), \\
& \gamma := -\frac{1}{2} \left( n^a \Delta \ell_a - \bar{m}^a \Delta m_a \right) = -\frac{1}{2} \left( n^a n^b \nabla_b \ell_a - \bar{m}^a n^b \nabla_b m_a \right), \\
& \beta := -\frac{1}{2} \left( n^a \delta \ell_a - \bar{m}^a \delta m_a \right) = -\frac{1}{2} \left( n^a m^b \nabla_b \ell_a - \bar{m}^a m^b \nabla_b m_a \right), \\
& \alpha := -\frac{1}{2} \left( n^a \bar{\delta} \ell_a - \bar{m}^a \bar{\delta} m_a \right) = -\frac{1}{2} \left( n^a \bar{m}^b \nabla_b \ell_a - \bar{m}^a \bar{m}^b \nabla_b m_a \right).
\end{aligned}
\end{equation}

Since the Riemann tensor can be decomposed into the Weyl tensor, Ricci tensor, and Ricci scalar, the curvature of spacetime can be described in the NP formalism using the following quantities:
\begin{equation}
\begin{aligned}
& \Psi_0=-C_{\mu \nu \rho \sigma} l^\mu m^\nu l^\rho m^\sigma, \\
& \Psi_1=-C_{\mu \nu \rho \sigma} l^\mu n^\nu l^\rho m^\sigma, \\
& \Psi_2=-C_{\mu \nu \rho \sigma} l^\mu m^\nu \bar{m}^\rho n^\sigma, \\
& \Psi_3=-C_{\mu \nu \rho \sigma} l^\mu n^\nu \bar{m}^\rho n^\sigma, \\
& \Psi_4=-C_{\mu \nu \rho \sigma} n^\mu \bar{m}^\nu n^\rho \bar{m}^\sigma,
\end{aligned}
\end{equation}
and
\begin{equation}
\begin{aligned}
\Phi_{00} &= -\frac{1}{2} R_{\mu \nu} l^\mu l^\nu, \\
\Phi_{22} &= -\frac{1}{2} R_{\mu \nu} n^\mu n^\nu, \\
\Phi_{02} &= -\frac{1}{2} R_{\mu \nu} m^\mu m^\nu, \\
\Phi_{20} &= -\frac{1}{2} R_{\mu \nu} \bar{m}^\mu \bar{m}^\nu, \\
\Phi_{11} &= -\frac{1}{4} R_{\mu \nu} \left( l^\mu n^\nu + m^\mu \bar{m}^\nu \right), \\
\Phi_{01} &= -\frac{1}{2} R_{\mu \nu} l^\mu m^\nu, \\
\Phi_{10} &= -\frac{1}{2} R_{\mu \nu} l^\mu \bar{m}^\nu, \\
\Lambda &= \frac{1}{12} R_{\mu \nu} \left( l^\mu n^\nu - m^\mu \bar{m}^\nu \right), \\
\Phi_{12} &= -\frac{1}{2} R_{\mu \nu} n^\mu m^\nu, \\
\Phi_{21} &= -\frac{1}{2} R_{\mu \nu} n^\mu \bar{m}^\nu.
\end{aligned}
\end{equation}

\section{Metric Reconstruction in Kerr Background}\label{procedure}
We provide a simplified version of the metric reconstruction method for Kerr BHs as presented in \cite{Loutrel:2020wbw}. For more details, please refer to \cite{Loutrel:2020wbw}.

For eq.(\ref{A9j}), since $\nu^{(1)}=0$ (gauge condition), we can get
\begin{equation}
(\Delta+\mu+\bar{\mu}+3 \gamma-\bar{\gamma}) \lambda^{(1)}=-\Psi_4^{(1)} .
\end{equation}
It is a transport equation. \(\lambda^{(1)}\) can be easily solved by usual numerical integration methods. 

Now consider eq.(\ref{c1a}). It can be found that the only non-zero metric component is $h_{\bar{m} \bar{m}}$. Since we have calculated \(\lambda^{(1)}\), one can get $h_{\bar{m} \bar{m}}$ by
\begin{equation}
[\Delta+2(\bar{\gamma}-\gamma)+\bar{\mu}-\mu] h_{\bar{m}\bar{m}}=-2 \lambda^{(1)} .
\end{equation}

For eq.(\ref{A10h}), since $\nu^{(1)}=0$, we can get an another transport equation for $\Psi_3^{(1)}$:
\begin{equation}
(\Delta+2 \gamma+4 \mu) \Psi_3^{(1)}=(\delta-\tau+4 \beta)\Psi_4^{(1)}+\mathcal{R}_h^{(1)} .
\end{equation}
Notice that $\mathcal{R}_h^{(1)}$ is related to energy-momentum tensor in GR.

The linearized eq.(\ref{A9i}) is
\begin{equation}
\begin{aligned}
(\Delta+\gamma-\bar{\gamma}) \pi^{(1)} & =-\mu(\pi+\bar{\tau})^{(1)}-\lambda^{(1)}(\bar{\pi}+\tau) \\
& -\Psi_3^{(1)}-\Phi_{21}^{(1)} .
\end{aligned}
\end{equation}
We can simplify it as
\begin{equation}
\begin{aligned}
(\Delta+\gamma-\bar{\gamma}) \pi^{(1)} & =\left(\frac{1}{2} \mu h_{\bar{m} \bar{m}}-\lambda^{(1)}\right)(\bar{\pi}+\tau) \\
& -\Psi_3^{(1)}-\Phi_{21}^{(1)},
\end{aligned}
\end{equation}
since there is a simple relation derived from eq.(\ref{C1l}) and (\ref{C1k}):
\begin{equation}
\pi^{(1)}+\bar{\tau}^{(1)}=-\frac{1}{2} h_{\bar{m} \bar{m}}(\bar{\pi}+\tau).
\end{equation}

With \(\pi^{(1)}\) given, one can get the equation based on eq.(\ref{C1l}):
\begin{equation}
(\Delta+\bar{\mu}-2 \bar{\gamma})^{(0)} h_{l \bar{m}}=-2 \pi^{(1)}-h_{\bar{m} \bar{m}} \tau^{(0)} .
\end{equation}

The final equation seems complicated, which is derived from eq.(\ref{A9f},\ref{A10g},\ref{C1g}):
\begin{equation}
\begin{aligned}
\left[ \frac{1}{4}(-\Delta + \gamma + \bar{\gamma}) \right. & \left. (-\Delta + 2 \bar{\gamma} + \mu - \bar{\mu}) + \frac{1}{2} \gamma (-\Delta + \gamma + \bar{\gamma}) \right. \\
& \left. - \frac{1}{2} \Delta \gamma \right] h_{l l} \\
= & \left[ -\frac{1}{4} (-\Delta + \gamma + \bar{\gamma}) (-\delta + 2 \bar{\alpha} - \bar{\pi} - 2 \tau) \right. \\
& \left. + \gamma (\bar{\pi} + \tau) \right] h_{l \bar{m}} \\
+ & \left[ -\frac{1}{4} (-\Delta + \gamma + \bar{\gamma}) (\bar{\delta} - 2 \alpha - 3 \pi - 2 \bar{\tau}) \right. \\
& \left. + \gamma (\pi + \bar{\tau}) \right] h_{l m} \\
+ & \left( \alpha^{(1)} - \frac{1}{2} \beta h_{\bar{m} \bar{m}} \right) (\bar{\pi} + \tau) \\
+ & \left( \beta^{(1)} - \frac{1}{2} \alpha h_{m m} \right) (\pi + \bar{\tau}) \\
+ & \pi \tau^{(1)} + \pi^{(1)} \tau + \Psi_2^{(1)}.
\end{aligned}
\end{equation}

\section{Quick Review of Modified Teukolsky Equation}\label{Mypaper}
For completeness, we briefly introduce the work presented in \cite{Guo:2024bqe}.

By applying the operators $(\delta+\bar{\pi}-\bar{\alpha}-3 \beta-4 \tau)$ and $(D-3 \epsilon+\bar{\epsilon}-4 \rho-\bar{\rho})$ to eq.(\ref{use}), we obtain:
\begin{equation}
\begin{aligned}
& (\delta + \bar{\pi} - \bar{\alpha} - 3 \beta - 4 \tau) \Big( (\bar{\delta} - 4 \alpha + \pi) \Psi_0^{(1)} \\
& \quad - (D - 4 \rho - 2 \epsilon) \Psi_1^{(1)} - 3 \kappa^{(1)} \Psi_2 + 2 \kappa^{(1)} \Phi_{11} \Big) \\
& - (D - 3 \epsilon + \bar{\epsilon} - 4 \rho - \bar{\rho}) \Big( (\Delta - 4 \gamma + \mu) \Psi_0^{(1)} \\
& \quad - (\delta - 4 \tau - 2 \beta) \Psi_1^{(1)} - 3 \sigma^{(1)} \Psi_2 - 2 \sigma^{(1)} \Phi_{11} \Big) \\
& = (\delta + \bar{\pi} - \bar{\alpha} - 3 \beta - 4 \tau) S_1- (D - 3 \epsilon + \bar{\epsilon} - 4 \rho - \bar{\rho}) S_2.
\end{aligned}
\end{equation}

Using the operator identity eq.(\ref{identity}), it can be simplified as 
\begin{equation}
\begin{aligned}
& (\delta+\bar{\pi}-\bar{\alpha}-3 \beta-4 \tau)\left((\bar{\delta}-4 \alpha+\pi) \Psi_0^{(1)}-3 \kappa^{(1)} \Psi_2+2 \kappa^{(1)} \Phi_{11}\right) \\
& -(D-3 \epsilon+\bar{\epsilon}-4 \rho-\bar{\rho})\left((\Delta-4 \gamma+\mu) \Psi_0^{(1)}-3 \sigma^{(1)} \Psi_2-2 \sigma^{(1)} \Phi_{11}\right) \\
& =(\delta+\bar{\pi}-\bar{\alpha}-3 \beta-4 \tau) S_1-(D-3 \epsilon+\bar{\epsilon}-4 \rho-\bar{\rho}) S_2.
\end{aligned}
\end{equation}

 Comparing with the GR case, the non-Ricci-flatness caused term appeared in left hand side is given by 
 \begin{equation}
 \begin{aligned}
(\delta+\bar{\pi}-\bar{\alpha}-3 \beta-4 \tau)\left(-3 \kappa^{(1)} \Psi_2+2 \kappa^{(1)} \Phi_{11}\right)\\
-(D-3 \epsilon+\bar{\epsilon}-4 \rho-\bar{\rho})\left(-3 \sigma^{(1)} \Psi_2-2 \sigma^{(1)} \Phi_{11}\right).
\end{aligned}
\label{ap1}
\end{equation}

Using the gauge condition 
\begin{equation}
\sigma^{(1)}=0, 
\end{equation}
Besides, $\delta$ contains only angular derivative. All of these properties lead eq.(\ref{ap1}) to 
\begin{equation}
\begin{aligned}
&(\delta+\bar{\pi}-\bar{\alpha}-3 \beta-4 \tau)\left(-3 \kappa^{(1)} \Psi_2+2 \kappa^{(1)} \Phi_{11}\right)\\
=&\left(-3 \Psi_2+2 \Phi_{11}\right)(\delta+\bar{\pi}-\bar{\alpha}-3 \beta-4 \tau) \kappa^{(1)} .
\end{aligned}
\end{equation}
 
Besides, the gauge condition $\sigma^{(1)}=0$ can simplify the third equation in eq.(\ref{use}) as:
\begin{equation}
-(\delta-\tau+\bar{\pi}-\bar{\alpha}-3 \beta) \kappa^{(1)}-\Psi_0^{(1)}=0 .
\end{equation}
And notice that in the spherical case, $\tau=0$. So
\begin{equation}
(\delta+\bar{\pi}-\bar{\alpha}-3 \beta-4 \tau) \kappa^{(1)}=(\delta-\tau+\bar{\pi}-\bar{\alpha}-3 \beta) \kappa^{(1)},
\end{equation}
the equations in this metric are once again decoupled, just
like the story in GR. We can get eq.(\ref{TeukEq}).
\end{appendices}

\bibliography{refs}

\end{document}